\documentclass[pra,twocolumn, preprintnumbers,showpacs,superscriptaddress]{revtex4} 
\usepackage{latexsym,epsfig,amssymb,amsfonts,amsmath,graphicx,bbm}
\usepackage[latin1]{inputenc}
\usepackage{exscale}

\usepackage{color}

\definecolor{dred}{rgb}{.8,0.2,.2}
\definecolor{ddred}{rgb}{.8,0.5,.5}
\definecolor{dblue}{rgb}{.2,0.2,.8}

\newcommand{\ii}{\mathrm{i}}
\newcommand{\LL}{\mathcal{L}}
\newcommand{\RR}{\mathcal{R}}

\newcommand{\dd}{ \mathrm{d}}

\newcommand{\half}{\mbox{$\textstyle \frac{1}{2}$}}
\newcommand{\ket}[1]{ | #1 \rangle}

\newcommand{\bra}[1]{ \langle #1 \,  |}

\newcommand{\od}[2]{\frac{\dd #1}{\dd #2}}

\newcommand{\eqr}[1]{Eq.~(\ref{#1})}
\newcommand{\fir}[1]{Fig.~\ref{#1}}

\newcommand{\an}[1]{\hat{#1}}
\newcommand{\cre}[1]{\hat{#1}^\dag}

\newcommand{\dt}[1]{\frac{\partial #1}{\partial t}}

\newcommand{\ee}{\textrm{e}}

\newcommand{\qq}{\mathbf{q}}

\newcommand{\rr}{\mathbf{r}}

\begin{document}

\title{Breathing oscillations of a trapped impurity in a Bose gas}

\author{T. H. Johnson}
\email{t.johnson1@physics.ox.ac.uk} 
\affiliation{Clarendon Laboratory, University of Oxford, Parks Road, Oxford OX1 3PU, United Kingdom}
\author{M. Bruderer}
\affiliation{Fachbereich Physik, Universit\"{a}t Konstanz, D-78457 Konstanz, Germany}
\author{Y. Cai}
\affiliation{Department of Mathematics, National University of Singapore, 119076, Singapore}
\author{S. R. Clark}
\affiliation{Centre for Quantum Technologies, National University of Singapore, 3 Science Drive 2, 117543, Singapore}
\affiliation{Clarendon Laboratory, University of Oxford, Parks Road, Oxford OX1 3PU, United Kingdom}
\affiliation{Keble College, University of Oxford, Parks Road, Oxford OX1 3PG, United Kingdom}
\author{W. Bao}
\affiliation{Department of Mathematics, National University of Singapore, 119076, Singapore}
\affiliation{Center for Computational Science and Engineering, National University of Singapore, 117543, Singapore}
\author{D. Jaksch}
\affiliation{Clarendon Laboratory, University of Oxford, Parks Road, Oxford OX1 3PU, United Kingdom}
\affiliation{Centre for Quantum Technologies, National University of Singapore, 3 Science Drive 2, 117543, Singapore}
\affiliation{Keble College, University of Oxford, Parks Road, Oxford OX1 3PG, United Kingdom}




\begin{abstract}
Motivated by a recent experiment [J. Catani {\em et al.}, arXiv:1106.0828v1 preprint, 2011], we study breathing oscillations in the width of a harmonically
trapped impurity interacting with a separately trapped Bose gas. We provide an intuitive physical picture of such dynamics at zero temperature, using a
time-dependent variational approach. In the Gross-Pitaevskii regime we obtain breathing oscillations whose amplitudes are suppressed by self trapping, due to interactions with the Bose gas. Introducing phonons in the Bose gas leads to the damping
of breathing oscillations and non-Markovian dynamics of the width of the impurity, the degree of which can be engineered through controllable parameters. Our results reproduce the main features of the impurity dynamics observed by Catani {\em et al.} despite experimental thermal effects, and are supported by simulations of the system in the Gross-Pitaevskii regime. Moreover, we predict novel effects at lower temperatures due to self-trapping and the inhomogeneity of the trapped Bose gas.
\end{abstract}

\maketitle


\section{Introduction}
The ability to trap and cool atoms of different species to ultra-low
temperatures has led to the realisation of various theoretical models in
which the intriguing physics of binary mixtures of spin hyperfine states or
different elements can be studied~\cite{Gunter2006,Ospelkaus2006,Widera2008,Lamporesi2010,Soltan-Panahi2011}. In particular, highly imbalanced
mixtures have made it possible to investigate the dynamics, interactions and
decoherence of single atoms, generally referred to as impurities, immersed
in a background atomic gas~\cite{Klein2007,Palzer2009,Bruderer2010,Goold2011,Johnson2011,Will2011}. As a prominent example, signatures of polaron
effects, caused by impurity-induced density fluctuations of the background
gas, have been investigated
theoretically~\cite{Bruderer2007,Tempere2009} and observed in
experiments~\cite{Catani2008,Gadway2010}.

More recently, Catani {\em et al.} created a harmonically trapped impurity
suspended in a separately trapped Bose
gas~\cite{Catani2011}. They studied the dynamics of the system
following a sudden lowering of the trap frequency of the impurity. Primarily, they observed breathing oscillations of
the width $\sigma$ of the impurity density distribution for various impurity-Bose gas interaction
strengths. Several features of the experiment were amenable to interpretation in terms of a quantum
Langevin equation in conjunction with a polaronic mass shift, however, we show that at even lower temperatures a different model is required to fully describe the dynamics of the system.

In this article, we develop a versatile analytical model describing the
experiment realised by Catani {\em et al.} at zero temperature. At first, our analysis
is based on a variational approach in the Gross-Pitaevskii (GP) regime.
We show that the impurity density distribution, which we describe by a Gaussian, has a width obeying a Newtonian equation of motion for a fictitious particle with position $\sigma$. The
potential governing this motion accounts for the quantum pressure of the
impurity, the inhomogeneity of the trapped Bose gas and the localised
distortion of this background gas induced by the impurity. The latter leads to a
strong confinement of the impurity, known as
self-trapping~\cite{Lee1992,Kalas2006,Bruderer2008,Luhmann2008}. Subsequently, we
extend our model by including excitations of the Bose gas in the form of
Bogoliubov phonons. A variational ansatz, which describes the bosons as a
product of coherent phonon states, allows us to track the evolution of the system, including the exchange of
energy between the impurity and the phonon bath. The dynamics turns out to be non-Markovian because of the back-action of
phonons created by the impurity, and we show that the timescale of memory effects can be varied by adjusting the trapping parameters, allowing for a comprehensive study of the transition between Markovian and non-Markovian dynamics.

Our results reproduce the main features observed in the experiment of
Catani~{\em et al.} despite neglecting finite temperature effects, but also demonstrate that cooling the system further should introduce novel effects. We obtain
an amplitude reduction of the breathing oscillation for large
impurity-Bose gas interactions. In particular, we observe a sudden quench of
the breathing amplitude once the impurity-Bose gas coupling substantially
exceeds the boson-boson interaction strength. Similar to the experiment, we find
that the breathing oscillations are damped due to the net dissipation of energy into the Bose gas. Moreover,
our model explains the different behaviour of attractive and repulsive
impurities as a consequence of the inhomogeneity of the trapped Bose gas.


\section{Model}\label{sec:analytics}
The specific system we consider consists of a single impurity of mass $m_a$ in a harmonic trap with frequency $\Omega_a$, and
identical bosons of mass $m_b$, trapped
by a separate external potential $v_b (\rr)$, whose densities weakly interact with strength $g$. The impurity
interacts with the Bose gas also via a density-density interaction of strength $\eta g$, where the dimensionless
parameter $\eta$ controls the relative strengths of interactions. As such, the many-body Hamiltonian for the combined
system is given by
 $\hat{H} = \hat{H}_a + \hat{H}_b + \hat{H}_{ab}$, where
\begin{subequations} \label{eq:Hamiltonian}
\begin{align}
\hat{H}_a &= \int \dd \rr \cre{\chi}  \left( - \frac{\hbar^2 \nabla^2}{2 m_a} + \frac{m_a}{2}\Omega_a^2 r^2 \right) \an{\chi} ,\\
\hat{H}_b &= \int \dd \rr \cre{\phi}  \left( - \frac{\hbar^2 \nabla^2}{2 m_b} + v_b + {\frac{g}{2}} \cre{\phi}  \an{\phi} \right)
\an{\phi} ,\\
\hat{H}_{ab} &= \eta g \int \dd \rr \cre{\chi}  \an{\chi}  \cre{\phi}  \an{\phi}  , \label{eq:interactionh}
\end{align}
\end{subequations}
describe the impurity, bosons and impurity-Bose gas coupling, respectively.
Here, $\an{\chi} (\rr,t)$ and $\an{\phi} (\rr,t)$ are the impurity and boson field operators. The single impurity has wavefunction $\chi(\rr,t)$.


\section{Breathing oscillations in the GP regime} \label{sec:TF}

In the GP approach, we replace $\an{\phi}(\rr,t)$ by the condensate mode of the
Bose gas $\varphi(\rr,t)$. For low temperatures and in a dilute regime it follows from Eqs. (\ref{eq:Hamiltonian}) that the impurity and condensate wavefunctions evolve according to
\begin{subequations} \label{eq:Hartree}
\begin{align}
i \hbar \dt \chi &= \left( - \frac{\hbar^2 \nabla^2}{2 m_a} + \frac{m_a}{2}\Omega_a^2 r^2 + \eta g | \varphi |^2 \right)\chi\,,\label{eq:abeforeTF}\\
i \hbar \dt \varphi &= \left( - \frac{\hbar^2 \nabla^2}{2 m_b} + v_b  + \eta g | \chi |^2 + g | \varphi |^2 \right)\varphi ,
\label{eq:bbeforeTF}
\end{align}
\end{subequations}
respectively, where $\varphi$ is normalised to the number of particles $N$ in the Bose
gas~\cite{Pitaevskii2003}.


\begin{figure*}[t]
\includegraphics[width=16.5cm]{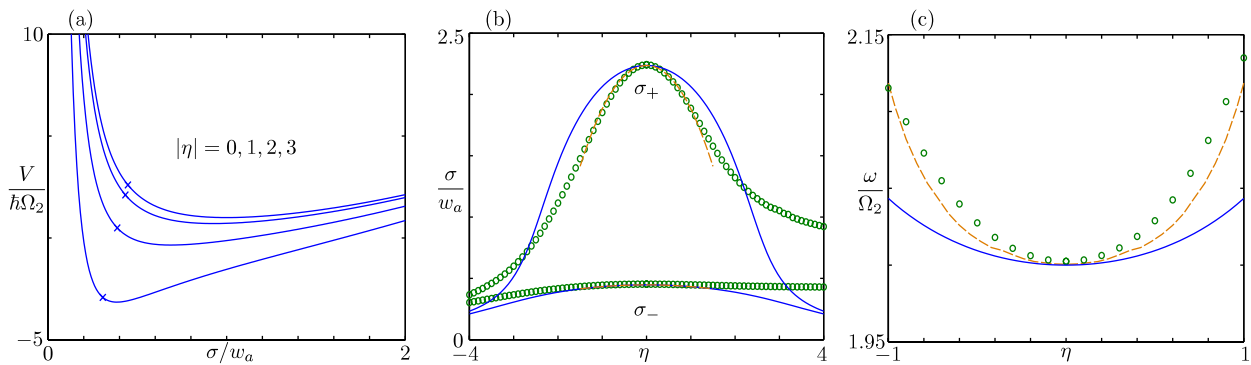}
\caption{\label{fig:TFapproach} {\em One-dimensional homogeneous Bose gas}.
(a) The potential $V$ governing the evolution of the width~$\sigma$ of an impurity
trapped with frequency $\Omega_2$ for various interactions strengths~$\eta$.
The crosses mark the width $\sigma$ at the minimum of $V$ for tight trapping
with $\Omega_1 = 5\Omega_2$.
(b) The maximum $\sigma_{+}$ and minimum $\sigma_{-}$ width during breathing oscillations
as a function of $\eta$. The analytical results of the GP approach predict that self-trapping strongly suppresses breathing oscillations (full blue). Additional suppression occurs in the Bogoliubov approach due to dissipation into phonon modes (dashed orange), matching the numerical results (green dots).
(c) The frequency of the breathing oscillations $\omega$ as a function of $\eta$.
The parameters are set to $m_b = 2 m_a$, $n_0 = \sqrt{m_b\Omega_2/\hbar}$
and $g = \hbar\Omega_2\sqrt{\hbar/m_b \Omega_2}$.}
\end{figure*}


We consider the condensate in the Thomas-Fermi regime, in which the kinetic
energy of the condensate, represented by the first term on the right hand
side of \eqr{eq:bbeforeTF}, is negligible. If the impurity and Bose gas are decoupled the condensate evolves as $\varphi_0(\rr,t) = \sqrt{n_0(\rr)} \mathrm{e}^{-i \mu_b t / \hbar}$, where the decoupled condensate density is given by $n_0( \rr) = [\mu_b - v_b(\rr)]/g$ for $\mu_b > v_b(\rr)$ and zero otherwise, and $\mu_b$ is the chemical potential. Returning to the case of non-zero impurity-boson coupling, we assume a similar form for the time-evolution of the condensate $\varphi(\rr,t) = \sqrt{n(\rr)} \mathrm{e}^{-i \mu_b t / \hbar}$. It then follows from
\eqr{eq:bbeforeTF} that, for attractive (repulsive) interactions, the
condensate density $n(\rr) =  n_0(\rr) - \eta |\chi(\rr)|^2$ is
enhanced (suppressed) from its decoupled value at the location of the
impurity. The self-consistency of our assumption regarding the
time-evolution of the condensate requires $n_0(\rr) \gg \eta |\chi(\rr)|^2$,
{\em i.e.}, sufficiently weak impurity-boson coupling.

As a consequence of these approximations,
\eqr{eq:abeforeTF} becomes the self-focussing non-linear Schr\"{o}dinger equation
\begin{equation} \label{eq:self-focussing}
i \hbar \dt \chi  = \left( - \frac{\hbar^2 \nabla^2}{2 m_a} + \eta g n_0 - \eta^2 g | \chi  |^2 + \frac{m_a}{2} \Omega_a^2 r^2 \right) \chi .
\end{equation}

We now focus on the important case of a spherically symmetric system in $d$
dimensions with the Bose gas trapped in the harmonic potential
$v_b( r) = \half m_b \Omega_b^2 r^2$ with frequency $\Omega_b$. The chemical potential in this case is
\begin{equation}
\mu_b = \hbar \Omega_b \left[ \frac{(2+d)\,\Gamma \left(1 + d/2\right)}{ 2\,(2 \pi)^{d/2}} \frac{N g}{\hbar
\Omega_b w^{d}_b} \right]^{2/(2+d)}, \nonumber
\end{equation}
where $w_b = \sqrt{ \hbar / m_b \Omega_b}$ is the width of the single particle
ground state and $\Gamma (p)= \int_0^\infty \dd x \ee^{-x} x^{p-1} $ is the Gamma
function. 

We solve \eqr{eq:self-focussing} within the variational ansatz
 \begin{equation}\label{eq:ansatz}
\chi(r , \sigma, \gamma) = \left( \pi \sigma^2  \right)^{-d/4} \mathrm{exp} \left(- r^2 / 2 \sigma^2  - \ii \gamma r^2 \right) ,
\end{equation}
with time-dependent width $\sigma$ and phase $\gamma$. This ansatz
was shown to describe a Bose gas in a harmonic trap accurately
(see~\cite{Castin2001} and references therein). We find equations of motion
for the Gaussian parameters by extremizing the action $S = \hspace{-1pt}
\int  \dd t\,\dd\rr \,\LL$ with the Lagrangian density
 \begin{align}
\LL = \chi^\ast \left[ i \hbar \dt{}  +
\frac{\hbar^2 \nabla^2}{2 m_a} - \eta g n_0 + \frac{\eta^2}{2} g | \chi |^2   -
 \frac{m_a}{2} \Omega_a^2 r^2 \right]\chi\,.  \nonumber \label{eq:lagrangianTF}
\end{align}
As a result, $\gamma = -m_a \dot{\sigma} / 2 \hbar \sigma$ and the width of
the impurity $\sigma$ obeys the equation of motion
\begin{equation}\label{eq:eomsigma}
    m_a\ddot{\sigma} = - \frac{\partial V(\sigma)}{\partial\sigma}
\end{equation}
with the potential $V = V_0 + V_{\mathrm{st}} + V_{\mathrm{inh}}$, where
\begin{subequations}
\begin{align}
   V_0(\sigma) & = \frac{\hbar^2}{2 m_a \sigma^2 } + \frac{m_a}{2}\Omega_a^2 \sigma^2\,, \\
   V_{\mathrm{st}}(\sigma) & = - \frac{\eta^2 g}{ d (2 \pi)^{d/2} \sigma^d} \vphantom{\int\limits_{\sum}^{\sum}}\,, \\
   V_{\mathrm{inh}}(\sigma) & =  \eta \mu_b \left[\frac{2}{d}\,\tilde{\Gamma} \left(
\frac{d}{2} , \frac{R^2}{\sigma^2} \right)  - \frac{\sigma^2}{R^2} \tilde{\Gamma} \left(1+\frac{d}{2}, \frac{R^2}{\sigma^2}  \right) \right]\,. \label{eq:Vinh}
\end{align}
\end{subequations}
Here, $R = \sqrt{2 \mu_b /m_b \Omega_b^2}$ is the Thomas-Fermi radius of the
condensate and $\tilde{\Gamma} (p,z) = [\Gamma(p) ]^{-1} \int_0^z \dd x \ee^{-x}
x^{p-1}  $ is the normalised lower incomplete Gamma function.

The two terms in $V_0$ describe the quantum pressure and the harmonic
trapping of the impurity. The contribution $V_{\mathrm{st}}$ is second order
in $\eta$ and represents self-trapping effects, which result from the
deformation of the condensate due to the interaction with the impurity. The
inhomogeneity in the condensate density due to its trapping gives rise to the
potential $V_{\mathrm{inh}}$ that is first order in $\eta$. It has a simple form: The expression in the square brackets of \eqr{eq:Vinh} is a monotonically decreasing function in $\sigma/R$, taking the value $2 / d$ at $\sigma/R = 0$ and asymptotically approaching zero as $\sigma/R \rightarrow \infty$. Its gradient is largest for widths $\sigma \sim R$.

We see that
\eqr{eq:eomsigma}, governing the evolution of the width $\sigma$, has just
the form of Newton's second law for the motion of a fictitious particle with
position $\sigma$ and mass $m_a$ in the potential $V(\sigma)$. This analogy
allows us to understand the behaviour of the width $\sigma$ as the impurity undergoes a procedure similar to that in the experiment by Catani {\em et
al.}~\cite{Catani2011}. In this procedure the impurity initially feels a tight trap
with frequency $\Omega_1$ and the coupled impurity-Bose gas system is cooled
to the ground state (although in the experiment such low temperatures were not achieved). At time $t=0$ the trap frequency is reduced to
$\Omega_2$ and the subsequent evolution of the width for $t \ge 0$ is
observed. In our model this corresponds to the width of the impurity sitting
at the minimum of $V$ with frequency $\Omega_a = \Omega_1$ for times $t<0$.
Then for $t \ge 0$ the width moves in the potential $V$ with lower trapping
frequency $\Omega_a = \Omega_2$.

Our approach, with $V=V_0$, exactly captures the correct evolution of the impurity
wavefunction for zero impurity-boson coupling. The square of the width oscillates harmonically
with frequency $2 \Omega_2$ according to
\begin{equation}\label{eq:exact}
\sigma^2 (t) =  \sigma_{+}^2 \mathrm{sin}^2 (\Omega_2 t) + \sigma_{-}^2 \mathrm{cos}^2 (\Omega_2 t) , \nonumber
\end{equation}
where $\sigma_{\pm}^2 =  w^2_a ( \Omega_1 / \Omega_2 )^{\pm 1}$ and $w_a = \sqrt{\hbar/m_a \Omega_2}$ is the width of the harmonic oscillator ground state.

\subsection{Homogeneous condensate}\label{sec:hom}


\begin{figure*}[t]
\includegraphics[width=16.5cm]{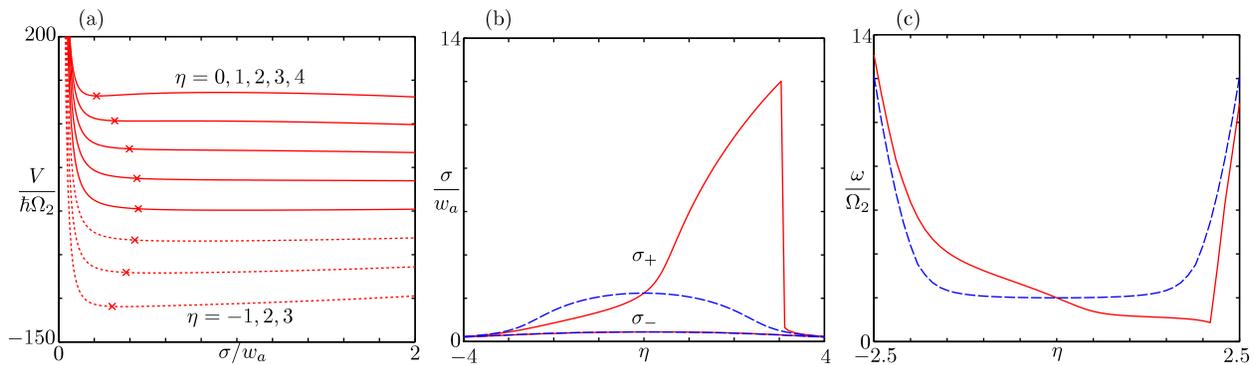}
\caption{\label{fig:TFapproach2} {\em One-dimensional Bose gas in a harmonic trap}.
(a)~The potential $V$ governing the evolution of the width $\sigma$ of
an impurity trapped with frequency $\Omega_2$ for positive (full) and negative
(dotted) interactions strengths $\eta$. The crosses mark the width $\sigma$ at
the minimum of $V$ for tight trapping with $\Omega_1 = 5\Omega_2$.
(b) The maximum $\sigma_{+}$ and minimum $\sigma_{-}$ width during breathing oscillations
as a function of $\eta$ for a trapped (full red) and homogeneous (dashed blue) Bose gas.
For a trapped Bose gas, breathing oscillations increase in amplitude with $\eta$ and at a critical value, here $\eta\approx3.1$, undergo
a sudden quench.
(c) The frequency of the breathing oscillations $\omega$ as a function of $\eta$.
The parameters are set to $m_b = 2 m_a$, $N = 200$, $\Omega_b =  \Omega_2 / \sqrt{2}$
and $g =\hbar \Omega_2\sqrt{\hbar/m_b\Omega_2}$.}
\end{figure*}


For non-zero impurity-boson coupling, we first consider the special case of a homogeneous condensate whose decoupled density $n_0$ is uniform, corresponding to the regime of a flat trap $R \gg \sigma$. In this case the potential $V_{\mathrm{inh}}$ is also uniform and can be neglected, leaving the potential $V$ symmetric with respect to attractive and repulsive interactions.

The self-trapping $V_{\mathrm{st}}$ imparts a
force on the width of the impurity towards smaller values of $\sigma$, shown
in \fir{fig:TFapproach}(a), and this has several effects. First, the minimum of the potential $V$ is shifted to smaller widths. Second, it shortens the distance between
the minimum of $V$ for $t <0$ and its minimum for $t \ge 0$. Third, the
curvature of potential $V$ near its minimum is increased. These effects
lead to an increased localisation of the impurity in the ground state at times
$t<0$. Further, at times $t\ge 0$ the amplitudes of the breathing
oscillations are reduced, along with an increase in the frequency of the
oscillations. This is shown in Figs.~\ref{fig:TFapproach}(b) and
\ref{fig:TFapproach}(c), where we have plotted the maximum and minimum
widths, and the frequency of the breathing oscillations, respectively. The
frequencies $\omega$ are found from the best fitting curve of the form $\sigma^2 (t) / w^2_a = A\mathrm{cos} (\omega t)+\sigma^2_0$ to the oscillations, where $A$ and $\sigma_0$ are fitting parameters.


\subsection{Condensate in a harmonic trap}

In the general case the decoupled condensate density $n_0$ depends on the
radial position $r$ and the potential $V$ is augmented by
$V_{\mathrm{inh}}$. The total potential $V$ resulting from a trapped
condensate is plotted in \fir{fig:TFapproach2}(a). Since the contribution of
$V_{\mathrm{inh}}$ is first-order in $\eta$ the symmetry between attractive
and repulsive impurity-boson coupling is broken. 

For repulsive coupling, the
force imparted by $V_{\mathrm{inh}}$ acts to increase the width of the
impurity, opposing self-trapping. Specifically, in the regime $\eta \lesssim 1$ the inhomogeneity of the Bose gas flattens the
potential $V$ near its minimum. In
contrast, for sufficiently large $\eta$ the self-trapping term dominates and
the potential develops a narrow trough at a small width, as for the
homogenous condensate. If interactions are attractive, $V_{\mathrm{inh}}$
represents a force on the width of the impurity towards smaller values of
$\sigma$ and enhances the self-trapping seen for the
homogeneous condensate.

The effects of $V_{\mathrm{inh}}$ on the dynamics of the impurity following
a sudden decrease in $\Omega_a$ are captured in Figs.~\ref{fig:TFapproach2}(b)
and \ref{fig:TFapproach2}(c). For weak repulsive interactions $\eta \lesssim 1$,
the effect of the flattening of the potential is to increase the amplitude
of oscillations and decrease the frequency. For strong repulsive
interactions $\eta\gg1$ the creation of a second narrow minimum in the
potential $V$ results in small amplitude and high frequency oscillations.
Our results predict a sudden transition between these two regimes, causing a sharp drop in the amplitude of
oscillations and an increase in frequency. For attractive impurities, the
effect of $V_{\mathrm{inh}}$ is to enhance the effects seen for a
homogeneous condensate. 


\section{Damping and memory effects}\label{sec:bog}

Up to now we have not included a mechanism for the transfer
of energy between the impurity and the Bose gas. We now show, for a homogeneous Bose gas, that the
oscillations of the impurity create Bogoliubov phonons and thereby lead to the exchange of energy with the Bose gas. Working in a volume $\mathcal{V}$ with periodic boundaries, we expand $\an{\phi}  = \varphi_0  + \delta \an{\phi}$. The decoupled mode $\varphi_0$ only describes the Bose gas in the ground state if $\eta = 0$; therefore, in this picture, interactions with the impurity generate excitations even in the ground state of the system. Terms in $\hat{H}$ higher than
second order in either the deformation $\delta \an{\phi}$ or the
coupling parameter $\eta$ are neglected~\cite{Ohberg1997,Bruderer2007}. Then, expressing the deformation $\delta
\an{\phi} = \sum_\qq [ u_\qq (\rr) \an{b}_\qq + v_\qq^\ast (\rr) \cre{b}_\qq
]$ in terms of Bogoliubov modes $\an{b}_\qq$ with momenta $\hbar \qq$, the Hamiltonian of the Bose gas
reads
\begin{equation}
    \hat{H}_b = E_0 + \sum_\qq \hbar \omega_\qq \cre{b}_\qq \an{b}_\qq \,. \nonumber
\end{equation}
Here, $E_0$ is the energy of the mode $\varphi_0$, $u_\qq $ and $v_\qq$ solve the Bogoliubov-de Gennes
equations, $\hbar \omega_\qq = \sqrt{\epsilon_\qq \left(
\epsilon_\qq +2g n_0 \right) }$ are the phonon energies, and $ \epsilon_\qq = \hbar^2 q^2/2 m_b$
are the free particle energies~\cite{Oosten2001}. Under the same approximations the interaction Hamiltonian simplifies to
\begin{equation}\label{eq:bogint}
    \hat{H}_{ab} = \eta g n_0 +\eta g \sum_{\qq\neq0} (\cre{b}_\qq +
\an{b}_\qq)\,f_\qq \,,
\end{equation}
with the coefficients
\begin{equation}
    f_\qq = \sqrt{\frac{n_0 \epsilon_\qq}{\mathcal{V} \hbar \omega_\qq}}\int \dd \rr
|\chi(\rr)|^2\mathrm{e}^{i \qq \cdot \rr}\,.  \nonumber \label{eq:f}
\end{equation}
The first and the second terms in
\eqr{eq:bogint} represent the interaction of the impurity with the condensate mode $\varphi_0$ and the Bogoliubov phonons $\an{b}_\qq$,
respectively. Moreover, the latter  describes the creation of Bogoliubov
phonons due to a classical driving force $f_\qq$. In our case, $f_\qq$
depends parametrically on the oscillating width $\sigma$ and thus the
driving force is approximately periodic.

We solve for both the ground state and evolution of the total system
variationally within the ansatz $\ket{\Psi} = \ket{ \sigma, \gamma} \otimes
\ket{ \{ \alpha_\qq \} }$. For the impurity, $\ket{ \sigma, \gamma}$
corresponds to the Gaussian ansatz in \eqr{eq:ansatz}. The Bose gas is
restricted to a product of coherent states $\ket{ \{ \alpha_\qq \} } =
\bigotimes_\qq \mathrm{e}^{-|\alpha_\qq |^2/2} \mathrm{e}^{\alpha_\qq
\cre{b}_\qq} \ket{0}$. This is known to describe bosonic modes coupled to a
classical field such as a density and in particular Bogoliubov modes coupled
to an impurity~\cite{Bruderer2007}.


\begin{figure}[t]
\includegraphics[width=8cm]{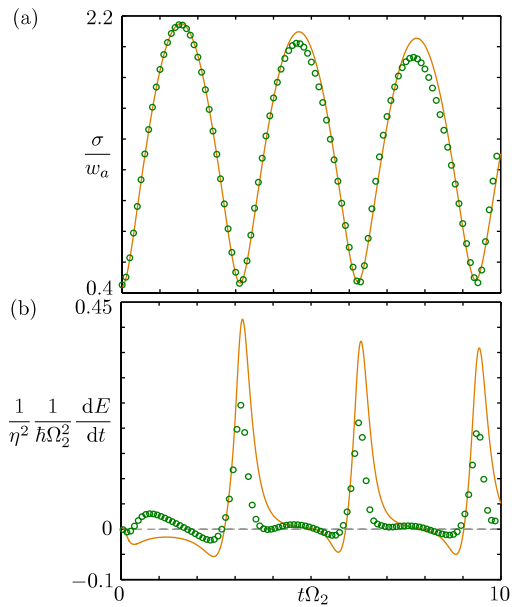}
\caption{\label{fig:bog} {\em Damping of oscillations due to loss of energy, for weak coupling}.
(a)~Damped oscillations of the width $\sigma$ of an impurity in a trap with frequency $\Omega_2$
obtained analytically (full orange) and numerically (green dots). The trap frequency for $t<0$ is
$\Omega_1 = 5\Omega_2$.
(b) The rate of energy exchange $\dot{E}$ between the impurity and the Bose gas.
Positive (negative) values correspond to energy gained (lost) by the Bose gas.
The parameters are set to $d=1$, $\eta = 0.5$, $m_b = 2 m_a$, $n_0 = \sqrt{m_b\Omega_2/\hbar}$
and $g = \hbar \Omega_2 \sqrt{\hbar / m_b \Omega_2 }$. The dashed line marks $\dot{E} = 0$.}
\end{figure}


We derive the equations of motion for the total system by extremizing the
action $S = \int\dd t\,L$ with the Lagrangian $L =
\bra{\Psi}\big(\ii\hbar\partial_t - \hat{H} \big) \ket{\Psi}$. Working in
the thermodynamic limit, where $\mathcal{V}^{-1}\sum_\qq \rightarrow 1/ (2
\pi)^d \int \dd \qq$, we evaluate the integrals analytically using the small
momentum approximation to the Bogoliubov dispersion relation $\omega_\qq = c
q $, with the speed of sound given by $c = \sqrt{g n_0 /m_b}$. This
approximation is accurate in the regime $q\ll m_bc/\hbar$. We found no qualitative differences when proceeding numerically without making this approximation.

Assuming the system is in the ground state at $t<0$, the resulting equations of motion describe the width of the impurity $\sigma$ moving in the potential $V_0$ augmented by a time-dependent force $F_{\mathrm{phon}}$, representing the interaction of the impurity with the phonon bath, given by
\begin{align}
& F_{\mathrm{phon}} (t)=  -  \frac{K \eta^2 g^2 n_0 \sigma(t)}{m_b c^2}   \Bigg\{ \frac{ \left[ \Sigma^2 (t,0) - c^2 t^2 \right]  \ee^{-\frac{c^2 t^2 }{ \Sigma^2 (t, 0)}
}}{
\Sigma^5 (t, 0 )} \label{eq:stforce}  \\ \nonumber
&+ c \int_0^t \dd t' \frac{ c (t-t')  \left[ 3 \Sigma^2 (t, t') - 2 c^2 (t - t')^2 \right] \ee^{- \frac{c^2 (t-t')^2 }{ \Sigma^2 (t, t')} }}{ \Sigma^7 (t, t' )} \Bigg\} , 
\end{align}
with compound width $\Sigma (t,t') = \left[ \sigma^2 (t) + \sigma^2 (t') \right]^{1/2}$ and constant $K = 2  d \sqrt{\pi} /  (4 \pi )^{d/2} \Gamma \left( 1 + d/2 \right))$.
The first and second terms in \eqr{eq:stforce} correspond to the
interaction of the impurity with phonons present in the initial state and
those created during the evolution, respectively. These interactions introduce non-Markovian effects with a memory that decays exponentially on a timescale $\sigma / c$. As a consequence, an experimentalist can, in principle, control the degree of non-Markovianity by altering $g$, $n_0$ and $w_a$. 

Due to the form of $F_{\mathrm{phon}}$, the equation of motion for
the width of the impurity $\sigma$ is an integro-differential equation,
which we solve using an iterative procedure: Initially, we evaluate
$F_{\mathrm{phon}}$ assuming a trajectory for $\sigma$ and integrate the equations of
motion, subsequently we use the resulting trajectory to evaluate $F_{\mathrm{phon}}$
in the next iteration. This procedure converges after a few iterations to a
self-consistent evolution for $\sigma$. The evolution of the width of the impurity after a sudden decrease in
$\Omega_a$ is plotted in \fir{fig:bog}(a). We find that the breathing
oscillations are damped out due to interactions with phonons.
In Figs.~\ref{fig:TFapproach}(b) and \ref{fig:TFapproach}(c) we plot the maximum and minimum widths, and frequencies as functions of coupling strength $\eta$. Frequencies of oscillations $\omega$ are found from the best
fitting curve $\sigma^2 (t) / w^2_a =  A e^{-\Lambda t}
\mathrm{cos} (\omega t) + C t + \sigma^2_0  $ to our data, where $A$, $B$, $C$, $\Lambda$ and $\sigma_0$ are fitting parameters. Dissipation before the first peak leads to a stronger suppression of maximum width than found when considering self-trapping only. We find that dissipation also results in a greater dependence of frequency on coupling. 

To confirm our analytical approach we have numerically solved the coupled time-dependent
GP Eqs.~(\ref{eq:Hartree}), the details of which can be found in the appendix. For small $|\eta|$, we find good agreement on the frequency of oscillations and the width at the first maximum, while the analytics underestimates the heights of later peaks, as exemplified in Figs.~\ref{fig:bog}(a). This leads to a close match of the numerically and analytically calculated maximum and minimum widths, and frequencies, as shown in Figs. \ref{fig:TFapproach}(b) and \ref{fig:TFapproach}(c). Also in these two figures, the numerics show that the symmetry between attractive and repulsive interactions is broken for large coupling; an attractive impurity can infinitely enhance the condensate density at the origin, while an infinitely repulsive impurity only depletes the density of the Bose liquid at its location to zero (the Moses effect)~\cite{Bruderer2008,Bible}.

We also calculate the energy deposited into the
Bose gas $E(t) = \sum_\qq \hbar \omega_\qq ( | \alpha_\qq (t) |^2 - |
\alpha_\qq (0) |^2 )$. The rate of energy lost by the impurity is
\begin{align}
& \od{E(t)}{t}  =   \frac{K \eta^2 g^2 n_0}{m_b c} \Bigg\{ - \frac{c t \ee^{-c^2 t^2 / \Sigma^2 (t, 0)
}}{
\Sigma^3 (t, 0 )} \nonumber \label{eq:energyloss}  \\ \nonumber
&+ c \int_0^t \dd t' \frac{ \left[ \Sigma^2 (t, t') - 2 c^2 (t - t')^2 \right] \ee^{-c^2 (t-t')^2 / \Sigma^2 (t, t') }}{ \Sigma^5 (t, t' )} \Bigg\} ,
\end{align}
which again is divided into two parts, representing interactions with phonons created at $t<0$ and $t \ge 0$, respectively. In \fir{fig:bog}(b) we plot this dissipation rate during the evolution shown in
\fir{fig:bog}(a). We find that most dissipation occurs when the width of the
impurity is near its minimum value and that $\dot{E}$ can take negative values (the impurity
can gain energy). It may, at first, appear at odds with the Landau criterion that dissipation is largest when the impurity density distribution is stationary. However, such arguments do not apply here, as in~\cite{Daley2004}, since the impurity is not simply moving at a constant velocity. The form of $\dot{E}$ also differs markedly from that predicted by
modelling the impurity as a damped harmonic oscillator, as done in~\cite{Catani2011}, whereby dissipation
is due to a friction force. The qualitative features of our analytic results are confirmed by numerical simulations, as shown in \fir{fig:bog}(b). Quantitative deviations arise from two approximations: We have neglected high order phonon effects and the impurity density does not always have a Gaussian form.


\section{Discussion and conclusion} \label{sec:conclusion}

Our results reproduce the main features of the experiment by Catani {\em et al.}~\cite{Catani2011} even though we assumed zero temperature. Both the finite temperature experiment and our theory exhibit breathing oscillations of the impurity after a
sudden decrease in trapping frequency. In particular, for $\eta =
0$ the oscillations observed in the experiment had a frequency and maximum to
minimum width ratio close to the exact zero temperature values. Also, in both experiment and theory, large impurity-Bose gas coupling suppresses the amplitudes of breathing oscillations and leads to damping. However, our zero temperature calculations predict a richer dependence of amplitude on coupling, including a sudden quench, resulting from the interplay of self-trapping and inhomogeneity induced by the Bose gas trapping. We also predict a dependence of oscillation frequency on the impurity-Bose gas coupling, indicating that the lack of a dependence observed by Catani {\em et al.} is a finite temperature effect. Our intuitive model explains differences between repulsive and attractive impurities, and highlights the way in which the non-Markovianity of the system arises and how it may be controlled.

In general, we expect the zero temperature approximation to be valid if the
thermal energy of the initial state $k_{\mathrm{B}} T$ is small compared to
the relevant energy scales of the system at $t <0$. In the experiment by
Catani {\em et al.}, prior to the equilibration the thermal energy of the Bose gas and the impurity were $350 \, \mathrm{nK}$
and $50 \, \mathrm{nK}$, respectively. We find that the typical energy scales of our system, $\hbar \Omega_1$ and the initial energy of the Bose gas, can be set to $125 \, \mathrm{nK}$ while retaining large oscillations $\Omega_1/ \Omega_2 \sim 10$ and the impurity width on the scale of micrometers. However, mechanisms governing the evolution
of the impurity at zero temperature, such as self-trapping, might still be
relevant up to the temperatures realised in the experiment.

Our flexible analytical framework allows for several natural extensions, {\em e.g.}, impurities interacting directly and via the Bose gas, in the same trap or displaced. Future work will analyse many-impurity effects in such systems.

\acknowledgements THJ thanks Vlatko Vedral and John Goold for interesting
discussions. MB acknowledges financial support from the Swiss National
Science Foundation through the project PBSKP2/130366. YC and WB acknowledge support by the Academic Research Fund of Ministry of Education of Singapore grant R-146-000-120-112. SRC and DJ thank the National Research Foundation and
the Ministry of Education of Singapore for support.


\appendix
\section{Numerical simulations}
We solve the coupled time-dependent
GP Eqs.~(\ref{eq:Hartree}) for a homogeneous Bose gas
numerically using time-splitting methods. We assume for numerical ease that the condensate is enclosed in an infinite spherical box with a radius $\RR$ that is large compared to the width $\sigma$ and the healing
length $\xi$. More precisely, we choose $\RR = 40 w_a$ and solve for $\varphi(r)$ and $\chi(r)$ on a spatial grid of $10^4$
points using timestep $10^{-3} \Omega_2^{-1}$. The ground states are calculated using the normalised gradient flow~\cite{Bao2004,Bao2004b,Bao2010} 
and the time-evolved states are calculated by applying a method combining the time-splitting technique to decouple the nonlinearity~\cite{Bao2003,Bao2004b} with the second order finite difference and Crank-Nicolson methods to discretise the spatial and temporal derivatives, respectively~\cite{Bao2006}.
To calculate the width of the impurity $\sigma$, we find the
best fitting parameters to $| \chi (r)|= B \mathrm{exp} (-r^2/ 2 \sigma^2)$.
Maximum and minimum widths are calculated from
this data.


\end{document}